\documentclass{aa}
\usepackage{graphicx}
\usepackage{txfonts}

\begin{document}

\newcommand{\Msun}{\ensuremath{\mathrm{M}_\odot}}
\newcommand{\Lsun}{\ensuremath{\mathrm{L}_\odot}}
\newcommand{\Rsun}{\ensuremath{\mathrm{R}_\odot}}
\newcommand{\Mdot}{\ensuremath{\dot{M}}}
\newcommand{\density}{\ensuremath{\mathrm{g/cm^3}}}
\newcommand{\arrow}{$\longrightarrow$}
\newcommand{\lsim}{\mathrel{\hbox{\rlap{\lower.55ex \hbox {$\sim$}}
 \kern-.3em \raise.4ex \hbox{$<$}}}}
\newcommand{\gsim}{\mathrel{\hbox{\rlap{\lower.55ex \hbox {$\sim$}}
 \kern-.3em \raise.4ex \hbox{$>$}}}}
\newcommand{\half}{\ensuremath{{\textstyle\frac{1}{2}}}}
\newcommand{\e}{\ensuremath{\mathrm{e}}}
\newcommand{\n}{\ensuremath{\mathrm{n}}}
\newcommand{\p}{\ensuremath{\mathrm{p}}}
\renewcommand{\H}[1]{\ensuremath{{^{#1}\mathrm{H}}}}
\newcommand{\D}{\ensuremath{\mathrm{D}}}
\newcommand{\He}[1]{\ensuremath{{^{#1}\mathrm{He}}}}
\newcommand{\Li}[1]{\ensuremath{{^{#1}\mathrm{Li}}}}
\newcommand{\Be}[1]{\ensuremath{{^{#1}\mathrm{Be}}}}
\newcommand{\B}[1]{\ensuremath{{^{#1}\mathrm{B}}}}
\newcommand{\C}[1]{\ensuremath{{^{#1}\mathrm{C}}}}
\newcommand{\N}[1]{\ensuremath{{^{#1}\mathrm{N}}}}
\renewcommand{\O}[1]{\ensuremath{{^{#1}\mathrm{O}}}}
\newcommand{\F}[1]{\ensuremath{{^{#1}\mathrm{F}}}}
\newcommand{\Na}[1]{\ensuremath{{^{#1}\mathrm{Na}}}}
\newcommand{\Ne}[1]{\ensuremath{{^{#1}\mathrm{Ne}}}}
\newcommand{\Mg}[1]{\ensuremath{{^{#1}\mathrm{Mg}}}}
\newcommand{\Fe}[1]{\ensuremath{{^{#1}\mathrm{Fe}}}}
\newcommand{\Ni}[1]{\ensuremath{{^{#1}\mathrm{Ni}}}}
\newcommand{\Co}[1]{\ensuremath{{^{#1}\mathrm{Co}}}}
\newcommand{\Al}[1]{\ensuremath{{^{#1}\mathrm{Al}}}}
\newcommand{\avg}[1]{\ensuremath{\langle#1\rangle}}
\newcommand{\rate}[1]{\ensuremath{r_\mathrm{#1}}}
\newcommand{\sv}[1]{\ensuremath{\langle\sigma v\rangle_\mathrm{#1}}}
\newcommand{\life}[2]{\ensuremath{\tau_\mathrm{#1}(\mathrm{#2})}}
\newcommand{\fpp}[1]{\ensuremath{f_\mathrm{pp#1}}}
                                                                                
\newcommand{\mwd}{$M_{\rm{WD}}$}
\newcommand{\msyr}{$\rm{M}_{\odot}/\rm{yr}$}
\newcommand{\kmpersec}{$\rm{km}\rm{s}^{-1}$}
\newcommand{\ctw}{$^{12}\rm{C}$}
\newcommand{\ost}{$^{16}\rm{O}$}
\newcommand{\rtrip}{$3\alpha$}
\newcommand{\rche}{$^{12}\rm{C}(\alpha,\gamma)^{16}\rm{O}$}

\authorrunning{S.-C. Yoon \& N.  Langer}
\titlerunning{Binary evolution toward a Chandrasekhar mass white dwarf }

\title{
The first binary star evolution model producing a Chandrasekhar mass white dwarf
}

\author{S.-C. Yoon \and N. Langer}

\institute{Astronomical Institute, Utrecht University, Princetonplein 5,
NL-3584 CC, Utrecht, The Netherlands}

\offprints {S.-C. Yoon, \email{S.C.Yoon@astro.uu.nl}}
\date{Received 25 September 2003 / Accepted 6 November 2003}


\abstract{Today, Type~Ia supernovae are essential tools for
cosmology, and recognized as major contributors to the chemical evolution
of galaxies. The construction of
detailed supernova progenitor models, however, 
was so far prevented by various physical and numerical difficulties
in simulating binary systems with an accreting white dwarf component,
e.g., unstable helium shell burning which may cause
significant expansion and mass loss.
Here, we present the first binary evolution calculation
which models both stellar components and the binary interaction
simultaneously, and where the white dwarf mass grows
up to the Chandrasekhar limit by mass accretion.
Our model starts with a $1.6$ $\rm{M_\odot}$ helium
star and a $1.0$  $\rm{M_\odot}$ CO white dwarf  in a
0.124~day orbit. 
Thermally unstable mass transfer
starts when the CO core of the helium star reaches $0.53 \rm{M_\odot}$,
with mass transfer rates of $1\cdots8 \times 10^{-6}$ $\rm{M_\odot/yr}$.
The white dwarf  burns the accreted helium steadily
until the white dwarf mass  has reached $\sim1.3$ $\rm{M_\odot}$ and
weak thermal pulses follow until
carbon ignites in the center when the white dwarf
reaches 1.37 $\rm{M_\odot}$.
Although the supernova production rate through this channel
is not well known, and this channel can not
be the only one as its progenitor life time is rather short 
($\sim 10^7 - 10^8 $ yr), our results indicate that helium star plus
white dwarf systems form a reliable route
for producing Type~Ia supernovae.
  
}
\maketitle


\section{Introduction}
Type Ia supernovae (SNe~Ia) are of particular importance in astrophysics:
They are the  major source for iron group elements in the universe
and are thus an essential contributor to 
the chemical evolution of galaxies (e.g. Renzini~\cite{Renzini}). 
And their light curve properties allow it to measure their distances
with an excellent accuracy even out to redshifts beyond $z=1$,
which makes them a powerful tool to determine the cosmological parameters 
(e.g. Hamuy et al.~\cite{Hamuy}; Branch~\cite{Branch}; Leibundgut~\cite{Leibundgut}). 
In particular, the recent suggestion of a non-zero cosmological constant
is in part based on SNe~Ia data (Perlmutter et al.~\cite{Perlmutter}; Riess et al.~\cite{Riess}).
An understanding of the progenitors of these supernovae is clearly
required as a basis for these fundamental astrophysical phenomena.

However, even though there seems no doubt that SNe~Ia are produced
by the thermonuclear explosion of a white dwarf,  
it is currently unclear in which kinds of binary systems such an event can occur
(Livio~\cite{Livio}).
Among the scenarios which have been put forward as possibilities,
the so called single degenerate scenario is currently favored,
where a CO white dwarf accretes mass from a non-degenerate companion
and thereby grows up to the Chandrasekhar mass 
(e.g. Hillebrandt \& Niemeyer~\cite{Hillebrandt}; Livio~\cite{Livio}).
However, hydrogen as well as helium accretion rates which allow
an increase of the CO white dwarf mass due to
shell burning are limited to narrow ranges 
(e.g. Nomoto~\cite{Nomoto}; Fujimoto~\cite{Fujimoto}; Iben \& Tutukov~\cite{Iben89}).
While hydrogen and helium shell sources are prone to
degeneracy effects and related thermonuclear instabilities
(e.g., nova explosions) at low accretion rates,
the helium shell source in accreting white dwarf models has been
found to be thermally unstable even for cases where the
electron degeneracy is negligible
(e.g. Cassisi et al.~\cite{Cassisi}; Langer et al.~\cite{Langer02}).
This affects in particular the potentially most frequent SN~Ia progenitor systems,
where a hydrogen rich star (main sequence star or red giant) is
considered as white dwarf companion 
(e.g. Li \& van den Heuvel~\cite{Li}; Hachisu et al.~\cite{Hachisu}; Langer et al.~\cite{Langer00}).
The hydrogen accretion rates in those systems 
which may allow steady hydrogen shell burning
are about a few $10^{-7}$ \msyr. 
With these rates, the subsequent helium shell burning
is usually found to be unstable
(Iben \& Tutukov~\cite{Iben89}; Cassisi et al.~\cite{Cassisi}; Kato \& Hachisu~\cite{Kato99}),
and no model sequences which cover the major part of the white dwarf accretion
phase could be constructed so far. 

In an effort towards overcoming this shortcoming, we consider here the
evolution of close helium star plus CO white dwarf systems.
In those, the white dwarf develops only a helium shell source,
thus avoiding complications involved in double shell source models
(e.g. Iben \& Tutukov~\cite{Iben89}).
Such systems form a predicted binary evolution channel
(Iben \& Tutukov~\cite{Iben94}), which is confirmed --- even though for lower
masses than considered here --- by Maxted et al. (\cite{Maxted}). 
Further evidence for the existence of
close helium star plus CO white dwarf systems comes from the
recent discovery of a helium nova
(Ashok \& Banerjee~\cite{Ashok}; Kato \& Hachisu~\cite{Kato03}).

Simplified binary evolution considerations provided us with an estimate
for the optimal initial parameters of our model system.
As a result, we embarked on
calculating the detailed evolution of a 1.6 \Msun{} helium star and a
1 \Msun{} CO white dwarf in a 0.124~d orbit.
We introduce our computational method and physical assumptions 
in Sect.~\ref{sec:method}. In Sect.~\ref{sec:results}, 
the evolution of the considered binary system is presented.
We discuss our results in Sect.~\ref{sec:discussion}.

\section{Numerical method and physical assumptions}\label{sec:method}

The numerical model has been computed
with a binary stellar evolution code which
computes the evolution of two binary components, and
the evolution of the mass transfer rate
and of the orbital separation simultaneously
through an implicit coupling scheme (Braun~\cite{Braun}).
We use Eggleton's (\cite{Eggleton}) approximation for the Roche lobe radius. 
Mass loss from the Roche lobe filling star through the first 
Lagrangian point is computed according to Ritter (\cite{Ritter}).
The orbital change due to stellar wind and mass transfer
is followed according to Podsiadlowski et al. (\cite{Pod}).
The specific angular momentum carried away by the stellar wind
is computed following Brookshaw \& Tavani (\cite{Brookshaw}).
Orbital angular momentum loss due to the gravitational 
wave radiation is also taken into account.
Opacities are taken from Iglesias \& Rogers (\cite{Iglesias}).
Mass loss due to a stellar wind is considered
as $\dot{M}= 10^{-2} RL/[GM(1-\Gamma)]$, with $\Gamma$
being the ratio of photospheric to Eddington luminosity.
This mass loss rate is based on dimensional
arguments and normalized for Wolf-Rayet stars (Langer~\cite{Langer89}).
However, the particular choice of the stellar wind 
mass loss formula does not affect our
final results significantly, given that the 
value of the mass loss rate is largely determined
by $\Gamma$, as
the luminosity of the white dwarf during the mass
accretion phase is close to the Eddington limit.
Effects of rotation are not considered here 
(cf. Langer et al.~\cite{Langer02}, Yoon et al.~\cite{Yoon}).
For more details about the code, see
Wellstein \& Langer (\cite{Wellstein99}) and Wellstein et al (\cite{Wellstein01}).

We start with a zero age main sequence helium star
of 1.6 \Msun{} and a 1.0 \Msun white dwarf, in 0.124 day orbit. 
The metallicity of the helium star is set to 0.02.
In order to avoid the numerical difficulty 
in following the initial strong helium shell flash in the white dwarf 
which might be induced soon after the onset of mass accretion,
the white dwarf is approximated by a point mass 
until the mass transfer rate from the helium star 
reaches about $10^{-6}$ \msyr\, (see Fig.~\ref{fig:mdot} in Sect.~\ref{sec:results}). 
To mimic the heating by the initial shell flash, 
we exchange the point mass by a hot  and bright
($\log L_{\rm s}/L_{\odot} = 4.175$) white dwarf model at
this time.
The initial central temperature and density of the white dwarf 
are $T_{\rm c}=1.8\times10^8$ K
and $2\times10^7$ \density{} respectively. Although the central temperature
may be much lower in reality when the mass accretion starts, 
the initial temperature structure is unimportant for the advanced evolution 
due to the self-heating of the white dwarfs (cf. Sect.~\ref{sec:results}).

The calculations presented here required high space and time resolution:
About 500$\,$000 binary models were computed, where both stars were resolved
into $\sim 1000$ mass shells.

\section{Results}\label{sec:results}

\begin{figure}[t]
\center
\resizebox{\hsize}{!}{\includegraphics{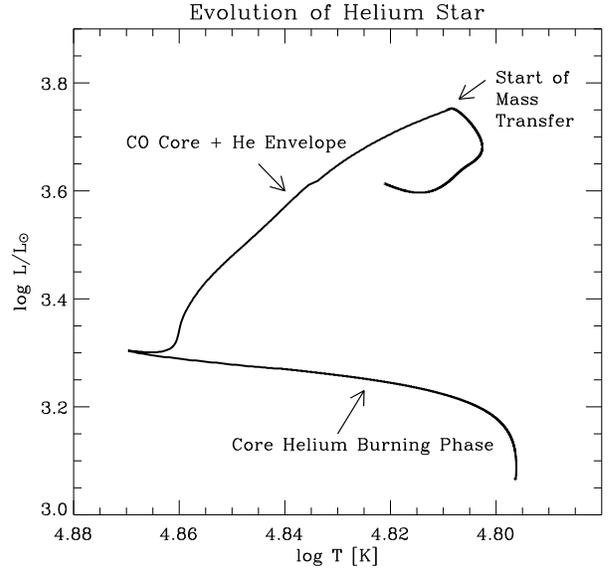}}
\caption{
Evolution of the 1.6 \Msun{} helium star in the HR diagram.
}\label{fig:hr1}
\end{figure}

\begin{figure}[t]
\center
\resizebox{\hsize}{!}{\includegraphics{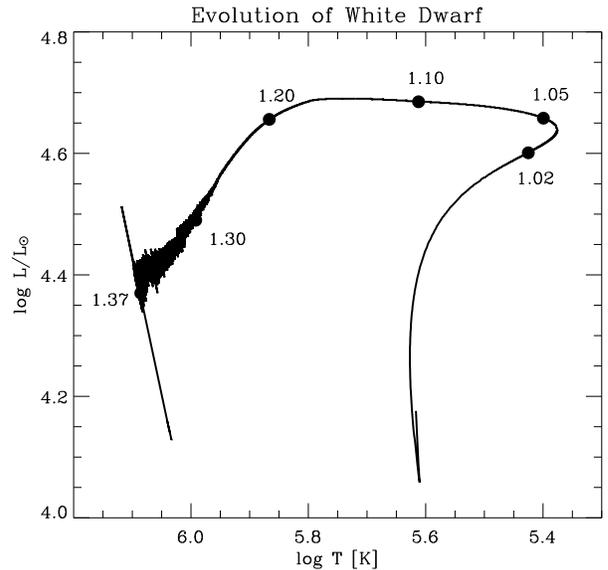}}
\caption{
Evolution of the CO white dwarf during the mass accretion
phase in the HR diagram.
The numbers at the filled circles denote the white dwarf mass
in the unit of solar mass
at the given points.
}\label{fig:hr2}
\end{figure}

\begin{figure}[t]
\center
\resizebox{\hsize}{!}{\includegraphics{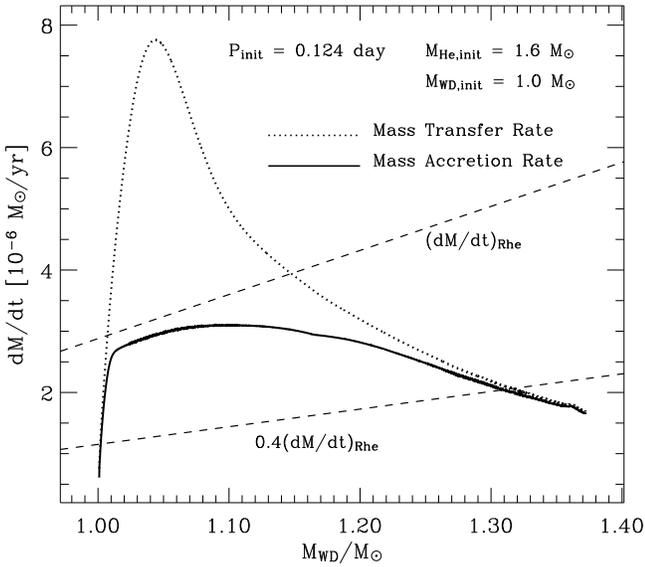}}
\caption{
Mass transfer rate (dotted line) and white dwarf mass
accretion rate (solid line) as function of the white dwarf mass,
for the considered binary model.
The upper dashed line denotes the critical mass accretion rate above which
the helium envelope is expected to expands to giant dimensions
(Nomoto~\cite{Nomoto}).
For accretion rates below the lower dashed line
helium shell burning is usually unstable.
}\label{fig:mdot}
\end{figure}

The evolution of our 1.6 \Msun{} helium star + 1 \Msun{} CO white dwarf
binary system proceeds as follows (see Figs.~\ref{fig:hr1} and~\ref{fig:hr2},
and Table~1).
The helium star undergoes the core helium burning for 4.28 million years.
After core helium exhaustion, the envelope of the helium star
expands and starts to fill its Roche lobe (Fig.~\ref{fig:hr1}) when
the radius of the helium star reaches 0.61 \Rsun.
At this point, the CO~core mass of the helium star reaches 0.53 \Msun{}
and mass transfer from the helium star commences.
Even though at this time the helium stars expands on the nuclear timescale
of helium shell burning, the mass transfer shrinks the orbit and thus
proceeds on the thermal timescale of the helium star. 
The resulting mass transfer rates are in the range $1\cdots8 \times 10^6$ \msyr, 
as shown by the dotted line in Fig.~\ref{fig:mdot}.

\begin{table}[t]
\begin{center}
\caption{Component masses, orbital period, and central density of both stars,
for five different times during the mass transfer, including 
the beginning of mass transfer (defines t=0), the time of minimum
orbital separation (t=73$\,$000$\,$yr), and the time of central carbon ignition
(t=148$\,$000$\,$yr).  }
\begin{tabular}{r c c c c c }
\hline \hline
t & $M_{\rm He}$ & $M_{\rm WD}$ & P & $\rho_{\rm c,He}$ & $\rho_{\rm c,WD}$ \\
$10^3\,$yr & \Msun & \Msun & h & $10^5\,$g$\,$cm$^{-3}$ & $10^8\,$g$\,$cm$^{-3}$\\ 
\hline
  0 & 1.60 & 1.00 & 2.97 & 1.31 & 0.20 \\
 35 & 1.39 & 1.10 & 2.70 & 1.53 & 0.51 \\
 73 & 1.25 & 1.21 & 2.65 & 1.86 & 1.45 \\
110 & 1.15 & 1.30 & 2.69 & 2.30 & 4.70 \\
148 & 1.08 & 1.37 & 2.78 & 2.92 & 23.5 \\
\hline
\end{tabular}
\end{center}
\end{table}

A part of the transferred matter is blown off by the
stellar wind at the white dwarf surface without being burned
into carbon and  oxygen, and thus without
affecting the thermal evolution of the CO core. 
The rest is transformed into CO in the helium burning shell
source and accreted into the CO core.
As the mass transfer rate from the helium star increases, 
the surface luminosity of the white 
dwarf reaches about 40\% of its Eddington luminosity (see Fig.~\ref{fig:hr2}).
A radiation driven wind from the white dwarf surface is induced,
with a peak of $4.8\times10^{-6}$ \msyr{} at
\mwd $\simeq$ 1.04 \Msun. 
As the mass transfer rate decreases, the stellar wind from the white 
dwarf becomes weaker. 
The upper dashed lines in Fig.~\ref{fig:mdot} denotes the critical mass accretion rate
above which the helium envelope is expected to expand to giant dimensions,
$\dot{M}_{\rm Rhe} = 7.2\times10^{-6} (M_{\rm CO}/M_\odot - 0.60) $ \msyr{}, 
based on white dwarf models computed with constant mass accretion rates
(Nomoto~\cite{Nomoto}).
The helium shell source is expected to be stable 
when $0.4\dot{M}_{\rm Rhe} <$ \Mdot{} $< \dot{M}_{\rm Rhe}$. 
The accretion rate in the white dwarf 
remains in the steady helium shell burning regime
until $M \simeq 1.3$ \Msun{} and the white dwarf mass grows
efficiently (Fig.~\ref{fig:hr2}). 
Thereafter, the white dwarf undergoes weak thermal pulses.

\begin{figure}[t]
\center
\resizebox{0.8\hsize}{!}{\includegraphics{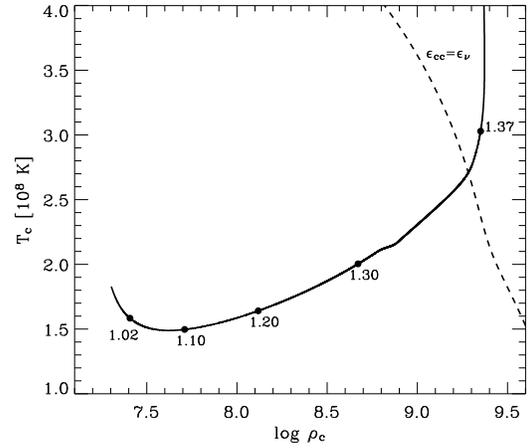}}
\caption{
Evolution of central density and temperature in the white dwarf during the mass accretion phase.
The dashed line gives the locus
where the energy generate rate due to the carbon burning equals to the neutrino cooling rate.
The numbers at the filled circles denote the white dwarf mass
in the unit of solar mass
at the given points.
}\label{fig:central}
\end{figure}

\begin{figure}[t]
\center
\resizebox{0.80\hsize}{!}{\includegraphics{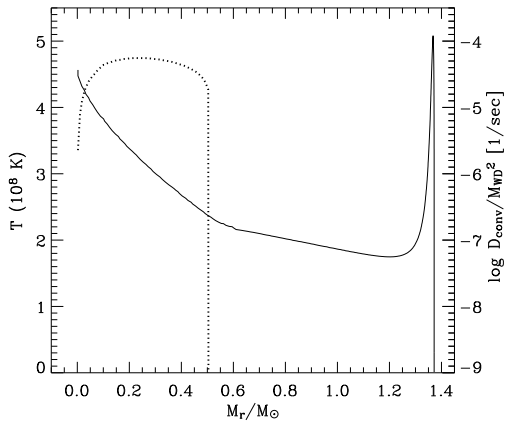}}
\caption{
Temperature as function of the mass coordinate in the last white dwarf
model. The dotted lines show the convective diffusion coefficient (in the unit of
${\rm g^2/s}$) divided by the square of the total mass of the white dwarf.
}\label{fig:final}
\end{figure}

During the mass accretion, 
the surface temperature of the white dwarf varies from $2.5\times10^5$ K to
$12 \times 10^5$ K
and  the luminosity reaches $2.5\cdots5 \times 10^4$ \Lsun{} (Fig.~\ref{fig:hr2}), 
with which the white dwarf will appear
as a super-soft X-ray source 
(e.g. Kahabka  \& van den Heuvel~\cite{Kahabka}; Greiner~\cite{Greiner1}).
The central temperature of the white dwarf decreases initially, mainly due to
neutrino emission.
When \mwd$\gtrsim$1.1 \Msun, the compressional heating
begins to dominate, and the central temperature of the white dwarf increases.
The dashed line in Fig.~\ref{fig:central} indicates
where the energy generation rate due to the carbon burning equals to the neutrino energy loss rate.
The white dwarf reaches this line at $\rho_{\rm c} = 1.92\times10^9$ ${\rm  g/cm^3}$
and  $T_{\rm c} = 2.71\times10^8$ K,
from where the central temperature increases rapidly.
About 1700 years thereafter, when $T_{\rm c} \simeq 2.9 \times 10^8$ K,
the central region becomes convectively unstable.
About 1800 years later, in our last model, 
the convection zone extends to $\simeq 0.5$ \Msun{}
when $T_{\rm c} = 4.6 \times 10^8$ K (Fig.~\ref{fig:final}).

The convective URCA process, which is not considered in our study, may 
affect the evolution during this stage (Paczynski~\cite{Paczynski}; Iben~\cite{Iben82}).
It is currently still debated whether it leads to heating
or to cooling of the convective core 
(e.g. Mochkovitch~\cite{Mochkovitch}; Stein et al.~\cite{Stein}).
Thus we have to consider the spatial extent and duration of the 
convective phase, as obtained from our models, as uncertain.
When the temperature exceeds $ 5\cdots8 \times 10^8$ K, 
the nuclear time scale becomes comparable to the convective time scale.
In this case, convection is not able to efficiently carry away the released
nuclear energy, and the mixing length theory becomes inadequate to describe
the convective energy transport. Therefore, we stop the calculation at this point;
see, however, H\"oflich \& Stein (\cite{Hoeflich}) for the further evolution.

In our last model, 
the helium star has a mass of 1.08 \Msun{} and a CO core mass of 0.64 \Msun. 
If the white dwarf explodes, the helium star may survive but
suffer stripping off small amounts of mass by the supernova ejecta 
(Marrietta et al.~\cite{Marrietta}). Its space velocity will be close to its final orbital
velocity of 330 km/s.
It will evolve into a massive CO white dwarf of $\sim1.0$ \Msun{} 
eventually.

\section{Discussion}\label{sec:discussion}

The model described above is (to our knowledge)
the first self-consistent binary evolution calculation
which leads to a Chandrasekhar-mass white dwarf.
Its relevance is in part the realistic construction of a 
supernova progenitor white dwarf model, but even more the fact that
it provides for the first time hard evidence of the functioning of
a SN~Ia progenitor channel: Our results leave little doubts that
some helium star plus white dwarf systems will in fact produce
a Type~Ia supernova.
  
The considered system may have evolved from a wide 8.0 \Msun{} giant star
+ 1.0 \Msun{} white dwarf system though a common envelope phase,
which in turn may be the result of two intermediate mass stars
in a close orbit.  
This indicates a system life time of the order of $\sim 10^7 - 10^8$ yr,
which is too short to explain SNe~Ia in elliptical galaxies.
Therefore, our results support the idea that at least two 
SN~Ia progenitor scenarios are realized in nature
(e.g., della Valle \& Livio~\cite{Valle}).
Iben \& Tutukov (\cite{Iben94}) estimated the potential SN~Ia production rate 
through white dwarf + helium giant binary systems to 
1.7$\times10^{-3}$ $\rm{yr^{-1}}$,  which may constitute 
a significant fraction of SNe~Ia observed in late type galaxies.

As previously mentioned, SNe~Ia progenitors of the kind considered here
will appear as a super-soft X-ray source (SSS).
We note that, since a wide range of orbital separations is possible
for helium star plus white dwarf systems (Iben \& Tutukov~\cite{Iben94}), 
this may explain SSSs with various orbital periods. 
For instance, 
short period systems such as RX J0537.7-7034 (3.5h, Greiner et al.~\cite{Greiner2})
and 1E0035.4-7230 (4.1h, Schmidtke et al.~\cite{Schmidtke})
can not be easily explained 
within the  canonical model which invokes hydrogen rich donor stars 
(e.g., Rappaport et al.~\cite{Rappaport}; Kahabka and van den Heuvel~\cite{Kahabka}),
except at low metallicity (Langer et al.~\cite{Langer00}).
Orbital periods of the binary system in the present study 
in the range $2.65 - 2.97$~h  
indicate that helium star plus white dwarf systems might be another natural 
possibility to explain short period SSSs.

\begin{acknowledgements}
We thank the referee for pointing out to us
the recent discovery of a helium nova. SCY is grateful to Philipp Podsiadlowski for
many helpful discussions during his visit to Oxford in February 2003, 
where this work was initiated. This research has been supported in part 
by the Netherlands Organization for
Scientific Research (NWO). 
\end{acknowledgements}


\begin{thebibliography}{}
\bibitem[2003]{Ashok} Ashok N.M., Banerjee D.P.K., 2003, A\&A 409, 1007
\bibitem[1998]{Branch} Branch D., 1998, ARA\&A, 36, 17
\bibitem[1997]{Braun} Braun H., 1997, Ph.D. Thesis, LMU M\"unchen
\bibitem[1993]{Brookshaw} Brookshaw L., Tavani M., 1993, ApJ 410, 719
\bibitem[1998]{Cassisi} Cassisi S., Iben I., Tornamb\'e A., 1998, ApJ 496, 376
\bibitem[1994]{Valle} della Valle M., Livio M., 1994, ApJ 423, L31
\bibitem[1983]{Eggleton} Eggleton P., 1983, ApJ 268, 368
\bibitem[1982]{Fujimoto} Fujimoto M.Y., 1982, ApJ 257, 767
\bibitem[2000]{Greiner1} Greiner J., 2000, New Astron. 5, 137
\bibitem[2000]{Greiner2} Greiner J., Orio M., Schwarz R., 2000, A\&A 355, 1041
\bibitem[1999]{Hachisu} Hachisu I., Kato M., Nomoto K., 1999, ApJ 522, 487
\bibitem[1996]{Hamuy} Hamuy M., Phillips M.M., Suntzeff N.B., et al. 1996, ApJ 519, 314
\bibitem[2000]{Hillebrandt} Hillebrandt W., Niemeyer J.C., 2000, ARA\&A 38, 191
\bibitem[2002]{Hoeflich} H\"oflich P., Stein J., 2002, ApJ 568, 779
\bibitem[1982]{Iben82} Iben I.Jr., 1982, ApJ 253, 248
\bibitem[1989]{Iben89} Iben I.Jr., Tutukov A.V., 1989, ApJ 342, 430 
\bibitem[1994]{Iben94} Iben I.Jr, Tutukov A.V., 1994, ApJ 431, 264 
\bibitem[1996]{Iglesias} Iglesias C.A., Rogers F.J., 1996, ApJ 464, 943
\bibitem[1997]{Kahabka} Kahabka, P, van den Heuvel, E.P.J., 1997, ARA\&A, 35, 69
\bibitem[1999]{Kato99} Kato M., Hachisu I., 1999, ApJ 513, L41
\bibitem[2003]{Kato03} Kato M., Hachisu I., 2003, to apper in ApJL, astro-ph/0310351
\bibitem[1989]{Langer89} Langer N., 1989, A\&A 210, 93
\bibitem[2000]{Langer00} Langer N., Deutschmann A., Wellstein S., H\"oflich P., 2000, A\&A 362, 1046
\bibitem[2002]{Langer02} Langer N., Yoon S.-C., Wellstein S., Scheithauer S. 2002, ASP Conference Proceedings, vol. 261, B.T. Gaensicke et al., eds., p.~252
\bibitem[2000]{Leibundgut} Leibundgut B., 2000, A\&AR 10, 179
\bibitem[1997]{Li} Li X.-D., van den Heuvel E.P.J., 1997, A\&A 322, L9
\bibitem[2001]{Livio} Livio 2001, In: Cosmic evolution, ed. E. Vangioni, R. Ferlet, M. Lemoine, New Jersey: World Sceintific
\bibitem[2000]{Marrietta} Marrietta E., Burrows A., Fryxell B. 2000, ApJS  218, 615
\bibitem[2000]{Maxted} Maxted P.F.L., March, T.R., North R.C. 2000, MNRAS, 317, L41
\bibitem[1996]{Mochkovitch} Mochkovitch R., 1996, A\&A 311, 152
\bibitem[1982]{Nomoto} Nomoto K., 1982, ApJ 253, 798
\bibitem[1972]{Paczynski} Paczy\'nski B., 1972, ApJ L11, 53
\bibitem[1999]{Perlmutter} Perlmutter S., Aldering G., Goldhaber G. et al., 1999, ApJ 517, 565 
\bibitem[1992]{Pod} Podsiadlowski Ph., Joss P.C., Hsu, J.J.L., 1992, ApJ 391, 246
\bibitem[1994]{Rappaport} Rappaport S.A., Di Stefano R., Smith M., 1994, ApJ 426, 692
\bibitem[1999]{Renzini} Renzini A., 1999, In: Chemical Evolution from Zero to High Redshift, Heidelberg, Springer Verlag, 185
\bibitem[2000]{Riess} Riess A.G., Filippenko A.V., Liu M.C., et al. 2000, ApJ 536, 62
\bibitem[1988]{Ritter} Ritter H., 1988, A\&A 202, 93
\bibitem[1996]{Schmidtke} Schmidtke P.C., Cowley A.P., McGrath T.K., Hutchings J.B., Crampto D., 1996, AJ 111, 788
\bibitem[1999]{Stein} Stein J., Barkat Z., Wheeler J.C., 1999, ApJ 523, 381
\bibitem[1999]{Wellstein99} Wellstein S., Langer N., 1999, A\&A 350, 148
\bibitem[2001]{Wellstein01} Wellstein S., Langer N., Braun H., 2001, A\&A 369, 939
\bibitem[2004]{Yoon} Yoon S.-C., Langer N., Scheithauer S., 2004, A\&A, in preparation

\end{thebibliography}
\end{document}